\documentclass[sigconf, nonacm=true]{acmart}

\usepackage[labelsep=period]{caption}
\usepackage{xcolor}
\usepackage[boxed]{algorithm2e}
\usepackage{lipsum}
\usepackage{microtype}

\def\mt#1{\mathrm{#1}}

\AtBeginDocument
{
	\providecommand\BibTeX{{\normalfont B\kern-0.5em{\scshape i\kern-0.25em b}\kern-0.8em\TeX}}
}

\SetAlgoCaptionSeparator{.}

\begin{document}

\title{\vspace*{-0.6cm}Metaheuristic macro scale traffic flow optimisation from urban movement data}

\author{Laurens Arp}
\authornote{Both authors contributed equally to this research.}
\email{{l.r.arp, d.van.vreumingen}@umail.leidenuniv.nl}
\author{Dyon van Vreumingen}
\authornotemark[1]
\affiliation
{%
  \institution{Leiden Institute for Advanced Computer Science}
  \streetaddress{Niels Bohrweg 1}
  \city{Leiden}
  \country{The Netherlands}
}
\author{Daniela Gawehns}
\email{{d.gawehns, m.baratchi}@liacs.leidenuniv.nl}
\author{Mitra Baratchi}
\affiliation
{%
  \institution{Leiden Institute for Advanced Computer Science}
  \streetaddress{Niels Bohrweg 1}
  \city{Leiden}
  \country{The Netherlands}
}

\renewcommand{\shortauthors}{L. Arp, D. van Vreumingen, D. Gawehns and M. Baratchi}

\begin{abstract}
How can urban movement data be exploited in order to improve the flow of traffic within a city? Movement data provides valuable information about routes and specific roads that people are likely to drive on. This allows us to pinpoint roads that occur in many routes and are thus sensitive to congestion. Redistributing some of the traffic to avoid unnecessary use of these roads could be a key factor in improving traffic flow.%\par
Many proposed approaches to combat congestion are either static or do not incorporate any movement data. In this work, we present a method to redistribute traffic through the introduction of externally imposed variable costs to each road segment, assuming that all drivers seek to drive the cheapest route. We use a metaheuristic optimisation approach to minimise total travel times by optimising a set of road-specific variable cost parameters, which are used as input for an objective function based on traffic flow theory. The optimisation scenario for the city centre of Tokyo considered in this paper was defined using public spatial road network data, and movement data acquired from Foursquare. %\par
Experimental results show that our proposed scenario has the potential to achieve a 62.6\% improvement of total travel time in Tokyo compared to that of a currently operational road network configuration, with no imposed variable costs.
\end{abstract}


\begin{teaserfigure}
%\vspace*{0.2cm}
%  \caption{Seattle Mariners at spring sraining, 2010.}
\end{teaserfigure}

\maketitle

\section{Introduction}
Even though extensive road networks have been developed to satisfy the high demand for vehicular transportation, overoccupancy of roads still occurs on a daily basis, causing traffic jams which hurt the environment, the economy and the drivers' moods. Finding a solution to traffic congestion is a challenging problem that has occupied many in the past century. After all, traffic dynamics are difficult to predict, due to complex fluctuations in traffic demand, both spatial and temporal. This makes it hard to devise a protocol for traffic flow redistribution that works well in varying conditions.\par
To date, various approaches have been proposed to alleviate congestion in some way \cite{KumarShukla2015}. However, these methods tend to be either static, data-independent protocols, micro-scale solutions (on the level of individual roads) or primarily driven by theoretical models. Our objective instead is to construct a dynamic, data-driven, macro-scale (road network level) approach to address traffic congestion. In this sense, dynamic means that a solution can be adapted to new traffic data with relative ease.\par
In this work, we propose a method for traffic redistribution fueled by metaheuristic optimisation, which we test on the case of the city centre of Tokyo. We seek to shift the traffic situation away from a state where each driver chooses the fastest, or shortest, route (thus causing congestion on roads that occur in many shortest routes), towards a system optimal equilibrium, as coined by Wardrop \cite{Wardrop1952}, where the \textit{total travel time} for all drivers is minimised. By introducing externally imposed variable costs (e.g. tolls, or any other financial or non-financial method a supervisory institution might deploy) on each road, we aim to discourage drivers from all taking the same congested roads. This approach asserts that, on average, each driver is willing to take the cheapest route from their point of departure to the destination, where the total costs to drive a route depend both on the distance travelled, through a spatial cost, and the imposed variable costs encountered along the route.\par
In order to make predictions of traffic flow and occurrence of congestion, we infer traffic demand from a data set of urban movements. A number of public traffic data sets, such as the Dutch NDW \cite{ndw}, report the traffic flow or density at certain points in time; however, while this gives a detailed picture of a local situation, it provides no information as to what routes drivers are following. Hence, such data is of little use when we wish to redistribute traffic by encouraging sensible alternative routes. For this reason, we use a data set of urban movements, provided by Foursquare as part of the \textit{Future cities challenge}, which allows for the inference of traffic level information needed for this research such as the origin and destination of movements \cite{fcc}.

\section{Related work}
The objective of combatting traffic congestion by altering road network setups has been addressed in a large body of work. The use of road pricing as a means to achieve this goal is a prevalent approach \cite{Keeler2002, Walters1961, Yang2017, Ye2012}. In this context, the marginal cost of congestion is a frequently employed measure to assess the optimal road pricing. A key difference between these papers and our work is that the previously proposed road pricing policies are fixed (e.g. charge a fee within a certain radius of the city centre) instead of dynamic, and do not follow from an optimisation procedure based on actual movement data. Approaches for optimising road networks and traffic flow from a different viewpoint, unrelated to road pricing policies, include metaheuristic optimisation of road improvements \cite{Gallo2010}, development of intelligent traffic light systems \cite{Pallavi2018}, optimisation of road graph architectures with evolutionary algorithms \cite{Schweitzer1997}, and prediction of optimal traffic flow through maximum-entropy methods \cite{Liu2008}.
%and even quantum implementations of traffic flow redistribution \cite{Neukart2017}.
A more exhaustive list of methods is provided by Kumar Shukla and Agrawal \cite{KumarShukla2015}. In this work, we explore the use of optimisation algorithms in proposing a dynamic pricing mechanism using actual movement data.

\section{Problem statement\label{par:prob}}
The problem of optimising traffic flow through adaptive road pricing is twofold. First, we must estimate traffic flow and congestion in a road network, which is the underlying cause of high total travel time when all drivers follow the cheapest routes from their origins to their destinations. By combining movement data describing the traffic demand in the network and the spatial road network data, traffic flow theory yields these estimations. The demand data is a set $D_M$ consisting of movements between venue locations within the road network. Each element of this set (indexed as $k$) is a tuple $(A_k, B_k, N_k)$ where $A_k$ and $B_k$ are elements of a venue data set $D_V$ containing spatial information about the venues, and $N_k$ is the recorded frequency of the specific movement from $A_k$ to $B_k$. Second, having found a method to express the total travel time as a function of the variable cost parameters, we aim to optimise the parameters for minimal total travel time. Our methods for addressing this optimisation problem are set out in detail in the next section.

\section{Methods}
\subsection{Traffic flow estimation}
\subsubsection{Road network and routing model}
The first step towards prediction and minimisation of congestion is to represent the physical road network as a planar graph $G$ that has road segments for edges, which may be traversed in order to travel from an origin to a destination. Specifically, the graph is a tuple $G=(V, E, S)$ with $V$ the node set, $E$ the edge set and $S$ the set of Haversine lengths %\cite{haversine}
of all edges. The node set $V$ contains intersections in the road network, as well as nodes for the origin and destination locations from $D_V$.\par
We then introduce, for each road segment $(i, j)\in E$ in the graph, a cost that a driver needs to pay to traverse this segment. The main part of this cost is a \textit{variable cost} $p_{ij}$. All variable cost parameters collectively form the variable cost vector $\mt P$ which we seek to optimise for minimum congestion. Next, we assign to each segment a \textit{spatial cost}, which is an immutable base cost for travelling from node $i$ to $j$ that is linearly dependent on the length $s_{ij}$ of the segment by a tunable factor $\beta_s$. Since the movement data is aggregated into frequency numbers, and is not provided on an individual level for anonymity reasons, we take $\beta_s$ to be equal for all drivers. Put together, the total cost for a driver to travel via a connected route of segments $R=\{(i, j)\}$ from some origin to a destination, is given by the sum of the individual segment costs occurring on the route:
\begin{equation}
{\rm cost}(R)=\sum_{(i, j)\in R}\beta_ss_{ij}+p_{ij}.
\end{equation}
For the development of traffic flow, we assert that all drivers are selfish and seek to drive the route which incurs the lowest total cost. These routes can be found using a weighted shortest path algorithm. Note that if $p_{ij}=0$ on all edges, each driver will drive the route of lowest spatial cost, which is exactly the shortest route. From the cheapest routes, which are jointed collections of segments, and the frequency numbers $N_k$, we can predict the vehicle count $n_{ij}$ on each segment, from which the degree of congestion is computed as set out in the following subsection.
\subsubsection{Congestion model}
In our congestion model, we assume that the flow of traffic on a road segment is fully described by a Greenshields fundamental traffic flow curve \cite{Greenshields1934}, %(see figure \ref{fig:funddiagram})
which is a widely used theoretical model for predicting traffic dynamics on a road segment. The used variables are flow $f$ (vehicles passing by per unit time) and density $\rho$ (vehicles present per unit length). In this model, there exists a \textit{maximum density} $\rho_m$ that the road can support, beyond which the total flow is zero. Furthermore, there is a \textit{critical density} $\rho_c$ at which the flow reaches its maximum value: $f(\rho_c)=f_m$. Naturally, $f(0)=0$, as no flow exists when no cars are present.\par
%\begin{figure}[t]
%  \centering
%  \includegraphics[width=0.65\linewidth]{Greenshields-traffic-flow-diagram}
%  \caption{A Greenshields fundamental traffic flow diagram. Below a critical density $\rho_c$, traffic flows freely and traffic flow $f$ increases with $\rho$. At densities beyond $\rho_c$, traffic becomes congested and traffic flow decreases.}
%  \Description{A Greenshields traffic flow diagram showing free traffic flow at low density and congestion at high density.}
%  \label{fig:funddiagram}
%\end{figure}\par
%Following the traffic flow diagram in figure \ref{fig:funddiagram}, we model the traffic flow as a quadratic function, with zero flow at $\rho>\rho_m$. Such a function is given by
A basic curve that fits this description is a concave quadratic function, with zero flow at $\rho>\rho_m$, which we define as
\begin{align}
f(\rho)=\frac{f_m}{\rho_c^2}\max\big(0, \rho[2\rho_c-\rho]\big),\label{vgl:f(rho)} 
\end{align}
where we note that, since $f$ is quadratic in $\rho$, $\rho_m=2\rho_c$. The maximum flow is directly related to the critical density; assuming that the traffic is able to drive at the maximum allowed speed $v_m$ when the density is at its critical point, we set $f_m=\rho_cv_m.$\par
From this density-flow dependence, we can extract the \textit{space mean speed} $v(\rho)$, the average speed of all vehicles on the road segment, as \cite{Greenshields1934}
\begin{align}
v(\rho)=\min\bigg(v_m, \frac{f(\rho)}\rho\bigg),\label{vgl:v(rho)} 
\end{align}
where again the maximum speed enters the relation, this time as a bound on the space mean speed. Finally, the \textit{space mean travel time} $t(\rho)$ on the segment, taken to have length $s$, is computed as
\begin{equation}
t(\rho)=\frac s{v(\rho)}.\label{vgl:t(rho)}
%This is a simple secondary school physics equation, no need for a reference here
\end{equation}
From the expected number of vehicles $n_{ij}$ on each road segment, we obtain the segment density as $\rho_{ij}=n_{ij}/s_{ij}$. For multi-lane roads, we divide this number by the number of lanes. By inserting the segment density into the density-time relation described above (eq. \ref{vgl:t(rho)}), we find the space mean time $t_{ij}$ spent on the segment. The collection of segment travel times then leads to the definition of the objective function for optimisation by a metaheuristic algorithm.
\subsection{Parameter optimisation}
\subsubsection{Objective function}
The objective function computes the measure ${\rm obj}(\mt P)$, which reflects the extent to which the system optimal equilibrium is reached by a variable cost configuration $\mt P$. This equilibrium occurs when the total travel times for each driver on their routes are minimal. As such, we define the objective function as the total space mean travel time over all routes driven on the road network. This total travel time can be conveniently expressed using the segment vehicle counts $n_{ij}$, which are directly dependent on $\mt P$:
\begin{equation}
{\rm obj}(\mt P)=\sum_{(i, j)\in E}n_{ij}(\mt P)\,t_{ij}(\mt P).%\,\Big/\!\sum_kN_k.
\label{vgl:obj}
\end{equation}
Note that each segment mean travel time $t_{ij}$ is also a function of $\mt P$ since it is dependent on the vehicle count $n_{ij}$. Algorithm \ref{alg:doelfunctie} shows, in pseudocode, the routine to compute the objective function.
\begin{algorithm}[!t]\small
\KwData{road graph $G=(V, E, S)$ with node set $V$ inferred from venue data $D_V$, $E$ and $S$ inferred from road network data;\newline
movement data $D_M$;\newline
spatial cost factor $\beta_s$}
\textbf{Input}: parameter vector $\mt P$\\
\KwResult{objective function value ${\rm obj}(\mt P)$}\vspace*{6pt}
initialise $n_{ij}\leftarrow0$, $t_{ij}\leftarrow0$, $\rho_{ij}\leftarrow0$ for all $(i, j)\in E$\\[6pt]
// Compute predicted vehicle counts $n_{ij}$ on each segment\\
\ForAll{origin-destination-freq. tuples $(A_k, B_k, N_k)$ in $D_M$}
{
	Find cheapest route $R_k$ from $A_k$ to $B_k$ according to $\beta_s$ and $\mt P$ using weighted 
	shortest-path algorithm\\
	\ForAll{$(i, j)\in R_k$}
	{
		$n_{ij}\leftarrow n_{ij}+N_k$
	}
}
// Find mean travel times $t_{ij}$ on each segment\\
\ForAll{$(i, j)\in E$}
{
	$\rho_{ij}\leftarrow n_{ij}/s_{ij}$\\
	$t_{ij}\leftarrow t(\rho_{ij})$\hspace*{13.7mm}
	(eqs. \ref{vgl:f(rho)}--\ref{vgl:t(rho)})
}
${\rm obj}(\mt P)\leftarrow \sum_{(i, j)\in E}n_{ij}\,t_{ij}$%\big/\!\sum_kN_k$
\hspace*{5mm}(eq. \ref{vgl:obj})\\
\textbf{return} ${\rm obj(\mt P)}$
\caption{Routine for computing the objective function for a given parameter vector $\mt P$.\vspace*{-0.4cm}}
\label{alg:doelfunctie}
\end{algorithm}
\subsubsection{Optimisation algorithms}
The purpose of the optimisation algorithms is to find an optimal variable cost configuration such that the value of the objective function, the total travel time, is minimised. In principle, any black-box metaheuristic optimisation algorithm could be used to search for local optima of variable cost configurations which might approximate a system optimal equilibrium. That said, for these purposes, algorithms which are robust for high-dimensional problems are preferred, as the number of parameters increases proportionally to the number of edges in the graph.\par
For our proof-of-concept implementation we use simulated annealing (SA) \cite{Kirkpatrick1983OptimizationBS}, which is a variation of hill climbing where worse solutions can get accepted depending on the algorithm's decreasing \textit{temperature} value, and a genetic algorithm (GA) adapted for continuous optimisation. %The GA had the general structure of a Genetic Algorithm, though uncanonically adjusted for real-valued optimisation and focusing on environmental selection.
For both algorithms each iteration contains 40 objective function evaluations; after one iteration, the GA updates its population, whereas SA resets its temperature value. Both algorithms use mutations generated using a normal distribution with zero mean and unit variance, at a mutation rate of 0.2 per parameter.
\section{Case study: Tokyo city centre\label{par:experiment}}
In order to test our traffic flow optimisation method, we applied it to movements inside the city centre of Tokyo (i.e. excluding the Greater Tokyo area). We briefly discuss the movements and road network used for the case study, and present experimental results.
\subsection{Movement data}
\begin{figure}[!t]
  \centering
  \includegraphics[width=0.65\linewidth]{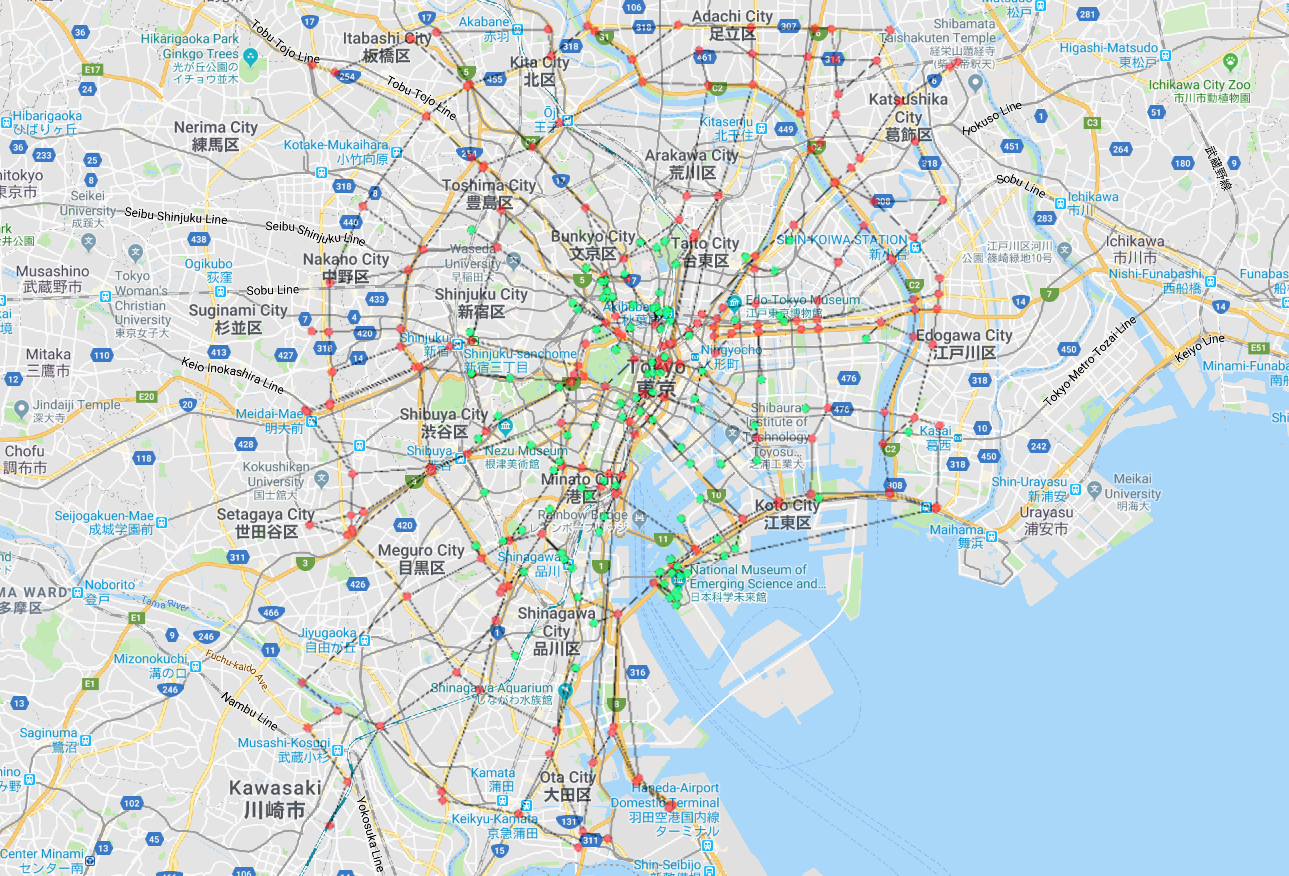}
  \caption{Road network graph example. Green nodes are venues, red nodes are intersections. Map source: Google Maps \cite{google}.\vspace*{-0.4cm}}
  \Description{Korte beschrijving}
  \label{fig:wegen}
\end{figure}\par
% Ik heb hem even hier gezet zodat hij op de vorige pagina komt

The movement data was provided by Foursquare as part of the \textit{Future cities challenge} \cite{fcc}; we selected only those parts related to the Tokyo city centre. The data contains a list of venues together with their GPS locations, forming the data set  $D_V$ introduced in section \ref{par:prob}, and a list of movements between the venues. The movement data contains movements of the same form as the tuples in $D_M$, but with additional indications of the month during which the movement occurred, and the time of the day (periods of 4 to 6 hours). We considered only movements made in the afternoon, and combined the frequencies $N_k$ for the same movement in different months into a single figure. Since the Foursquare dataset does not cover the entirety of vehicular movements inside Tokyo (and, in fact, also includes other types of movements such as subway, biking and walking trips), we viewed the frequencies as ratios rather than absolute numbers, asserting the law of large numbers for sufficient accuracy. We then normalised the frequencies to numbers that the road graph we constructed (see next subsection) could support.\par
As a last modification, we selected the venues occurring in the 100 most popular (i.e. frequent) routes, and clustered all other venues together with their nearest neighbour (in terms of Haversine distance) from the set of most popular venues. The routes were clustered accordingly, going between venue clusters instead of individual venues. This was done in order to substantially reduce the number of routes and therewith the computational complexity of the problem.
\subsection{Road network data}
The road network data used is based on the Asia shapefile provided by the Earthdata \textit{Global roads open access data set} \cite{eosdis}. It contains information on the road networks of the entirety of Asia with a variable resolution; for the city Tokyo, its resolution is well suited for the algorithm. The road data is translated into a graph representation by finding intersections between lines, and turning these into intersection nodes. The lines themselves are used to create edges between intersections. Venue nodes are created by identifying the location of the venues from the venue data set, which are connected to the nearest intersection node.
\subsection{Experimental results}
The improvement progress for both optimisation algorithms is shown in figures 3 and 4. For comparison, we ran the objective function once with all variable costs set to 0 to obtain the default flow of the road network. This configuration had an objective value of around 416 million hours spent on the road network. It should be noted that the objective function values are higher than what would be a realistic amount of time spent on roads. As the theoretical traffic flow models do not take relief methods into account for fully congested roads, the speed on those roads in considered to be zero. In these cases, we use an arbitrary low value to represent the speed on the road, resulting in somewhat unrealistic amounts of hours.
%It should be noted that the objective function values are higher than what would be a realistic amount of time spent on roads, due to the nature of the theoretical traffic flow models the objective function was based on when traffic is fully congested.

% TODO: ik snap dat je bovenstaande hebt gecomment, maar we zullen toch ergens moeten noemen dat dit geen realistische aantallen zijn voor 6 uur binnenstad tokyo... tenzij de gehele bevolking van japan elke zes uur een uur in het verkeer in tokyo doorbrengt (iirc zitten ze ergens in de honderden miljoenen qua bevolkingsaantal) 

\begin{figure}[t]
  \centering
  \includegraphics[width=0.65\linewidth]{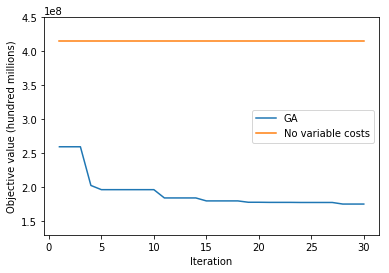}
  \caption{Objective function value plotted to the number of GA iterations. Values are the total time spent on the roads over 6 hours in the afternoon.\vspace*{-0.1cm}}
  \Description{Korte beschrijving}
  \label{fig:ga_progress}
\end{figure}\par

\begin{figure}[t]
  \centering
  \includegraphics[width=0.65\linewidth]{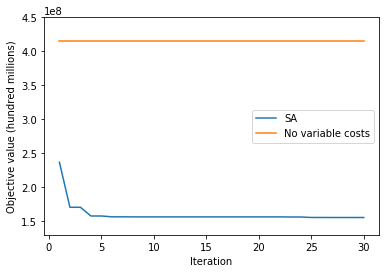}
  \caption{Objective function value plotted with the number of SA iterations. Values are the total time spent on the roads over 6 hours in the afternoon.\vspace*{-0.3cm}}
  \Description{Korte beschrijving}
  \label{fig:sa_progress}
\end{figure}\par

Both the GA and SA were able to find solutions which substantially improved traffic flow over the network. After 30 iterations, the lowest fitness value of the GA was around 175 million hours (an improvement of 57.9\%), and the lowest fitness value of SA was around 155 million hours (an improvement of 62.6\%). Effectively, this means the solution found by SA was able to reduce the total amount of hours spent over a period of 6 hours in the afternoon by 261 million, due to having found a good distribution of variable costs for each road segment. Although the results of the GA were not as good as those of SA, the GA was still successful in improving the unoptimised configuration, which supports the view that any optimisation algorithm could be used for these purposes. \\
Interestingly, even the first iterations of the algorithms showed values which were an improvement compared to the default traffic flow. This is likely due to the nature of the shortest path algorithm when taking only distance into account, which sends many cars over the same roads unnecessarily. Though this type of behaviour was the basic premise allowing us to optimise, given the large variety of available, very similar roads in the network, any deviation from the over-use of single roads caused by the random initialisation of variable costs would result in better traffic flow. From such a randomly initialised state, both algorithms then improved the solutions such that the more well-suited alternatives were used.\par
The results show that the algorithms were effective in finding solutions improving traffic flow over the default setting of no variable costs. That said, there is no guarantee the optima the algorithms converged to were global optima, nor that the convergence was as fast as possible. Future work could include more thorough exploration of optimisation algorithms and their parameter settings.

\section{Conclusion}
We have shown that we can successfully address traffic congestion by redistributing traffic through imposing of variable road segment costs, and optimising this cost configuration using metaheuristic algorithms. The best variable costs configuration was found by a simulated annealing routine, improving upon the total travel time corresponding to a configuration with zero variable costs by 62.6\%. Both simulated annealing and a genetic algorithm were effective at optimising solutions.\par
Though the practical implementation of the variable costs may be another non-trivial problem to address first, the positive results show that, at least conceptually, this method could result in improved traffic flow when applied in practice. Consequently, cities may enjoy shorter travel times, better accessibility, cleaner air and, not unimportantly, improved drivers' moods. 

\begin{acks}
We would like to extend special thanks to Foursquare and Netmob for the movement data and to Giorgos Kyrziridis for his support.
%We would like to extend special thanks to Foursquare and Netmob for organising the \textit{Future cities challenge} and providing the necessary data for this work. In addition, we would like to thank Giorgos Kyziridis for his assistance in realising this research.
\end{acks}

\bibliographystyle{ACM-Reference-Format}
\bibliography{biblio}

\appendix
%\section{Leuke appendix, zo nodig}
%\subsection{Oreo's}
%\lipsum[1]
%\subsection{Custardcakejes}
%\lipsum[2]

\end{document}